\documentstyle[aps,prd]{revtex}
\begin{document}
\thispagestyle{empty}
{\baselineskip0pt
\leftline{\large\baselineskip16pt\sl\vbox to0pt{\hbox{Department of Physics}
               \hbox{Kyoto University}\vss}}
\rightline{\large\baselineskip16pt\rm\vbox to20pt{\hbox{KUNS-1352}
           \hbox{July 1995}
\vss}}%
}
\vskip3cm
\begin{center}
{\large {\bf On a Quasi-Local Energy Outside the Cosmological Horizon}}
\end{center}
\begin{center}
 {\large Ken-ichi Nakao}\\
{\em Department of Physics,~Kyoto University} \\
{\em Kyoto 606-01,~Japan}
\end{center}
\begin{abstract}

We investigate a quasi-local energy naturally introduced by Kodama's
prescription for a spherically symmetric space-time with a
positive cosmological constant $\Lambda$.
We find that this quasi-local energy is well behaved inside
a cosmological horizon.
However, when there is a scalar field with a long enough Compton
wavelength, the quasi-local energy diverges in the course
of its time evolution outside the cosmological horizon.
This means that the quasi-local energy has a meaning only inside the
cosmological horizon.


\end{abstract}


In a spherically symmetric space-time, an energy (or equivalently
a mass in Einstein's theory) is well defined in a quasi-local manner since
there are not degrees of gravitational radiations.
Its reasonable expression was derived first by Misner and
Sharp\cite{Ref:MS} and later Kodama obtained the same one
but by a different procedure in which a conserved current is
introduced\cite{Ref:KODAMA}.
This Misner-Sharp energy is deeply related to the local
structure (e.g. the existence and
location of the trapped region and the type of singularities, etc)
and the dynamics of the space-time.
Further, the Misner-Sharp energy agrees with the well defined
global energy, i.e., the Bondi-Sachs
energy at the null infinity\cite{Ref:BONDI,Ref:SACHS}
and with the Arnowitt-Deser-Misner (ADM) energy at the spatial
infinity\cite{Ref:ADM} in the case of the asymptotically flat
space-time. Recently Hayward has extensively discussed it and shown
its geometrical and physical properties and meaning\cite{Ref:HAYI}.

In this paper, we shall investigate the behavior of the
quasi-local energy associated with matter fields in a
spherically symmetric space-time with a positive cosmological constant
$\Lambda$. The quasi-local energy in the spherically symmetric
space-time with $\Lambda$ is naturally introduced
by Kodama's prescription.
As will be shown, this quasi-local energy agrees with
the Misner-Sharp one in the case of $\Lambda=0$ and
further agrees with the Abbott-Deser (AD) energy\cite{Ref:AD,Ref:NSM}
at the ``spacelike infinity" for the asymptotically de Sitter
space-time, which is a conserved Killing energy in a global sense.

One of differences between the asymptotically flat space-time and
asymptotically de Sitter one is the existence of the
cosmological horizon defined as an outermost spacelike
closed two-surface enclosing the symmetric center (we
assume the spherical symmetry)
such that the expansion of the future directed
ingoing null orthogonal to the surface vanishes.
The quasi-local energy in the asymptotically de Sitter space-time
is well behaved inside the cosmological horizon like as the
Misner-Sharp one.
However, here it is worthy to note that the AD energy is not
bounded below in contrast with the ADM energy\cite{Ref:NSM}.
The negative contribution to the AD energy
comes from the distribution of matter fields outside the
cosmological horizon\cite{Ref:AD,Ref:NSM} and
correspondingly,
when we consider a scalar field with a long enough Compton
wavelength, the quasi-local energy diverges
negatively outside the cosmological horizon in the course
of its time evolution.
Hence, the quasi-local energy has a meaning only
inside the cosmological horizon in such situations.
In this paper, we follow the notation of
Misner, Thorne and Wheeler\cite{Ref:MTW} for the sign convention
of metric, etc and adopt the geometrical unit, i.e.,
$c=\hbar=G=1$.

The space-time with a cosmological
constant $\Lambda>0$ which we consider here
is governed by the Einstein equation,
\begin{equation}
G_{\mu\nu}
=8\pi \Bigl(T_{\mu\nu}-{\Lambda \over 8\pi}g_{\mu\nu}\Bigr).
\end{equation}
As already mentioned, we focus on the spherically symmetric
space-time and without loss of generality, the metric of it
is written as
\begin{equation}
ds^{2}=-\alpha^{2}(\eta,r)d\eta^{2}+A^2(\eta,r)
 dr^{2}+R^{2}(\eta,r)
(d\theta^{2}+\sin^{2}\theta d\varphi^{2}).
\end{equation}

Following Kodama\cite{Ref:KODAMA},
we shall first introduce a preferred ``time" vector
$K^{\mu}$. This Kodama's time vector is given by
\begin{equation}
 K^{\eta}=-{1 \over \alpha A}{\partial R\over \partial r},~~~~
 K^{r}=+{1 \over \alpha A}{\partial R\over \partial \eta},~~~~
{\rm and}~~~~
 K^{\theta}=0=K^{\varphi}.
\end{equation}
Then the current $S^{\mu}$ is defined by contracting $K^{\mu}$
with the right hand side of Eq.(1) as
\begin{equation}
S^{\mu}=S_{(M)}^{\mu}+S_{(\Lambda)}^{\mu},
\end{equation}
where
\begin{equation}
S_{(M)}^{\mu}=\alpha AR^{2}T^{\mu}_{\nu}K^{\nu}, ~~~~{\rm and}~~~~
S_{(\Lambda)}^{\mu}=-{1 \over 8\pi}\alpha AR^{2}\Lambda K^{\mu}.
\end{equation}
As shown by Kodama, the current $S^{\mu}$ satisfies
the conservation law $\partial S^{\mu}/\partial x^{\mu}=0$.
Here it should be noted that $S_{(M)}^{\mu}$ and
$S_{(\Lambda)}^{\mu}$ also satisfy
the conservation law separately, i.e.,
\begin{equation}
{\partial \over \partial x^{\mu}} S_{(M)}^{\mu}=0~~~{\rm and}~~~
{\partial \over \partial x^{\mu}}S_{(\Lambda)}^{\mu}=0.
\end{equation}

Since the conservation law (6) is
defined by the ordinary derivative, we can rewrite those into
integral forms separately.
We shall focus on the integral form only of the
first one in Eq.(6),
\begin{equation}
{d \over d\eta}M(r,\eta)=-4\pi S_{(M)}^{r}(r,\eta),
\end{equation}
where
\begin{equation}
M(r,\eta)\equiv 4\pi \int_{0}^{r}S_{(M)}^{\eta}(x,\eta) dx.
\end{equation}
Eqs.(7) and (8) mean that
$M$ can be regarded as a kind of a quasi-local energy associated
with matter fields.

{}From the Einstein equation, we obtain
\begin{equation}
S_{(M)}^{\eta}={1\over8\pi}\alpha AR^{2}
\Bigl(G^{\eta}_{\nu}+\Lambda \delta^{\eta}_{\nu}\Bigr)K^{\nu}.
\end{equation}
Using the above equation, we get another useful expression for
$M$ as
\begin{equation}
M={1 \over 2}R\Bigl[1+{1 \over2}R^2
\Bigl(\rho_{+}\rho_{-}-{2 \over 3}\Lambda
\Bigr)\Bigr],
\end{equation}
where $\rho_{+}$ and $\rho_{-}$
is the expansions of future directed outgoing and ingoing null,
respectively and are given by
\begin{equation}
\rho_{\pm}=-{1\over\sqrt{2}}
\Bigl(K_{T}\mp{2\over AR}{\partial R\over \partial r}\Bigr),
\end{equation}
where
\begin{equation}
K_{T}\equiv -{2\over \alpha}{\partial \over \partial \eta}\ln R.
\end{equation}

Hayward has shown that in the case of $\Lambda=0$,
the above quasi-local energy $M$ in the untrapped region,
$\rho_{+}>0$ and $\rho_{-}<0$, is non-decreasing in any outgoing
spatial or null direction if the matter fields
satisfy the dominant energy condition\cite{Ref:HAYI}.
Following Hayward's proof completely, we can see
that same is true in the case of $\Lambda >0$, if
the energy-momentum tensor except for $\Lambda$, i.e.,
$T_{\mu\nu}$ in Eq.(1), satisfies the dominant energy
condition (see Appendix A).
This means that the quasi-local energy $M$ is well behaved quantity
within the untrapped region.
If the region including the origin $r=0$ is untrapped,
the quasi-local energy $M$ is non-negative within this region
since a regular matter distribution implies $M(\eta,r=0)=0$.

Further, we should note that on a marginal surface $\rho_{+}=0$
or $\rho_{-}=0$, Eq.(10) becomes
\begin{equation}
1-{2M \over R}-{\Lambda \over 3}R^{2}=0.
\end{equation}
The above relation is same as that satisfied on event horizons
in the Schwarzschild-de Sitter space-time if the quasi-local energy
$M$ is identified with the mass parameter of that
space-time\cite{Ref:CARTER}. From Eq.(13), we see
that the quasi-local energy is related to
the existence and location of the trapped region, marginal
surfaces (apparent horizon and cosmological horizon) by an
expected manner.

Here let's consider the asymptotically de Sitter space-time which
is here naively defined as, for $r\rightarrow \infty$,
\begin{equation}
\alpha,~A~{\rm and}~R/r \longrightarrow a(\eta) + {\cal O} (r^{-1}),
\end{equation}
and
\begin{equation}
K_{T} \longrightarrow  -2H+{\cal O} (r^{-3})
\end{equation}
where $a(\eta)$ is the scale factor of the de Sitter space-time,
\begin{equation}
a(\eta)=-{1\over H\eta}~~~~~~{\rm with}~~~-\infty < \eta < 0,
\end{equation}
and $H=\sqrt{\Lambda/3}$ is the inverse of the de Sitter radius.
In this case, for $r\rightarrow\infty$, the quasi-local energy
$M$ agrees with the AD energy
$M_{AD}$\cite{Ref:AD,Ref:NSM}
which is the conserved Killing
energy in the asymptotically de Sitter space-time and
defined by
\begin{equation}
M_{AD}=-\lim_{r \rightarrow \infty}\Bigl[
r{\partial R \over \partial r}-rA
+{H\over2}R^{3}(K_{T}-2H)\Bigr].
\end{equation}
This fact shows that our definition of the quasi-local energy
is reasonable.

However it should be noted that in the case of $\Lambda>0$
there is a cosmological horizon.
Outside of the cosmological horizon is
not untrapped and therefore the quasi-local energy $M$
need not be non-decreasing in outgoing null or spatial direction.
Here we shall consider a scalar field and show that the
quasi-local energy $M$ negatively diverges outside the cosmological horizon.
We assume that the back reaction of the scalar field
to the space-time geometry is negligible and hence
we can treat a scalar field in the de
Sitter space-time, $\alpha =A =R/r=a(\eta)$.

Then the equation of motion for a scalar field is given by
\begin{eqnarray}
{\partial \over \partial \eta}k_{\phi}&=&-{H\eta \over r^{2}}
{\partial \over \partial r}\Bigl(r^{2}{\partial \over \partial r}
\phi\Bigr)+{3 \over \eta}k_{\phi}
+{1 \over H\eta}(m^{2}+12H^{2}\xi)\phi,\\
{\partial \over \partial \eta}\phi&=&-{1 \over H\eta}k_{\phi},
\end{eqnarray}
where $m$ is a mass of the scalar field and $\xi$
is a constant which determines the coupling between
the scalar field and the gravity\cite{Ref:BIR}.

First, for simplicity, we consider the case of $m=\xi=0$.
Differentiating Eq.(18) with respect to $\eta$ and using Eq.(19),
we obtain
\begin{equation}
\Bigl({\partial^{2} \over \partial \eta^{2}}
-{\partial^{2} \over \partial r^{2}})\Bigl({r \over \eta^{2}}k_{\phi}
\Bigr)=0,
\end{equation}
and this can be easily solved. The solution is obtained
as
\begin{eqnarray}
k_{\phi}&=&{\eta^{2} \over r}[Q(\eta+r)-Q(\eta-r)], \\
\phi&=&-{1\over Hr}\int_{\eta_{0}}^{\eta}d\omega\omega
[Q(\omega+r)-Q(\omega-r)]+\phi_{0}(r),
\end{eqnarray}
where $\eta_{0}$ corresponds to an initial conformal time,
$Q(x)$ is an arbitrary functions and $\phi_{0}(r)$
is also an arbitrary function but should
satisfy
\begin{equation}
{1 \over r^{2}}{d \over dr}\Bigl(r^{2}{d\phi_{0} \over dr}\Bigr)
=-{1 \over H\eta}\Bigl({\partial k_{\phi} \over \partial\eta}
-{3\over \eta}k_{\phi} \Bigr)|_{\eta=\eta_{0}}.
\end{equation}

Here it should be noted that as $\eta \rightarrow 0$,
\begin{eqnarray}
k_{\phi} &\propto& a^{-2} \rightarrow 0, \\
\phi&\rightarrow&-{1\over Hr}\int_{\eta_{0}}^{0}d\omega\omega
[Q(\omega+r)-Q(\omega-r)]+\phi_{0}(r),
\end{eqnarray}
and, in general, the right hand side of Eq.(25) does not vanish.
The above behavior is well-known fact that the massless scalar
field does not decay but is frozen in the de Sitter
space-time\cite{Ref:BIR}.
Of course, if $\phi$ and $k_{\phi}$ are small enough initially, these remains
small
for all time and therefore our assumption of no back reaction is
justified.

The time evolution of the quasi-local energy $M$ is determined
by Eq.(7).  From this equation, the rate of change of the quasi-local
energy $M$ is proportional to the flux $S^{r}_{(M)}$
which is given by
\begin{equation}
S^{r}_{(M)}={1 \over 2}a^{3}r^{2}\Bigl[
\Bigl(k_{\phi}-{1\over a}{\partial \phi \over \partial r}\Bigr)^{2}
+(Har-1)\Bigl\{k_{\phi}^{2}+{1\over a^{2}}
\Bigl({\partial\phi\over \partial r}\Bigr)^{2}
\Bigr\}\Bigr]
\end{equation}

The cosmological horizon of the de Sitter space-time is located at
$ar=H^{-1}$. From Eq.(26), we find that always $S^{r}_{(M)} \geq0$ outside
the cosmological horizon and hence from Eq.(7) the
quasi-local energy $M$ monotonically decreases with time outside
the cosmological horizon. Furthermore, for $\eta\rightarrow 0$
at $r=$constant,
\begin{equation}
S^{r}_{(M)} \longrightarrow {r^{3}
\over 2}Ha^{2}\Bigl({\partial\phi\over \partial r}\Bigr)^{2}
\propto a^{2},
\end{equation}
and hence form Eq.(7), we obtain
\begin{equation}
M \longrightarrow -2\pi ar^{3}
\Bigl({\partial\phi\over \partial r}\Bigr)^{2}
\propto a.
\end{equation}
The above equation shows that the quasi-local energy
negatively diverges as long as
$\partial\phi/\partial r|_{\eta=0}\neq0$.

The energy divergence outside the cosmological horizon
comes from the non-decaying outgoing flux $S_{(M)}^{r}$
due to the asymptotic fall-off behavior of the
massless scalar field $\phi \rightarrow $non-zero.
Even though the scalar field has a mass, if the mass is small enough,
the quasi-local energy $M$ outside the cosmological horizon diverges.
When $m \neq 0$ and $\xi \neq 0$,
the asymptotic behavior of the scalar field
is given by\cite{Ref:BIR}
\begin{equation}
\phi \longrightarrow f(r)a^{-{3\over2}+\sqrt{{9\over4}-{m^{2} \over
H^{2}}-12\xi}},
\end{equation}
where $f(r)$ is some function determined by the initial
configuration of the scalar field.
Hence if the scalar field minimally couples with gravity,
i.e., $\xi=0$, and the Compton
wavelength of the scalar field $m^{-1}$
is larger than $2H^{-1}/\sqrt{5}$,
the quasi-local energy $M$ diverges
for $\eta \rightarrow 0$
since $S^{r}_{(M)}\propto a^{-1-\epsilon}$, where
$\epsilon \geq 0$.

The energy divergence outside the cosmological horizon
means that the information inside the cosmological horizon
can not be obtained by the quasi-local energy $M$ defined outside
the cosmological horizon. Therefore, for the asymptotically
de Sitter space-time, the
quasi-local energy $M$ within the cosmological horizon seems to be
more important to specify
the structure of the
space-time than the global conserved quantity like as
the AD energy in such situations.

I would like to thank H. Sato for his encouraging
discussion and S.A. Hayward for his important suggestion and
discussion.
This work was partly supported by  Grant-in-Aid for Encouragement
of Young Scientists (No.07740354) from the
Ministry of Education, Science and Culture.

\appendix
\section*{A}

We shall see that the quasi-local energy $M$ is non-decreasing
in the untrapped region. The proof is completely same as
Hayward\cite{Ref:HAYI} and
we shall write the metric in the double-null form as
\begin{equation}
ds^{2}=-2e^{-f}d\zeta_{+}\zeta_{-}+R^{2}(d\theta^{2}
+\sin^{2}\theta d\varphi^{2}).
\end{equation}
The above form is obtained by taking $\alpha=A=e^{-f/2}$
and $\zeta_{\pm}=(\eta \pm r)/\sqrt{2}$ in Eq.(2).
Here we shall write the quasi-local energy as
\begin{equation}
M(r,\eta)={1 \over 2}R\Bigl[1+{1 \over2}R^2
\Bigl(e^{-f}\vartheta_{+}\vartheta_{-}-{2 \over 3}\Lambda
\Bigr)\Bigr],
\end{equation}
where $\vartheta_{\pm}=2R^{-1}\partial_{\pm}R=e^{f/2}\rho_{\pm}$
and $\partial_{\pm}$ means the derivative with respect to
$\zeta_{\pm}$.
Without loss of generality, the components of the
energy-momentum tensor in the spherically symmetric space-time is
written as
\begin{equation}
T_{\pm\pm}=\psi_{\pm},~~T_{+-}=2\varepsilon~~
{\rm and}~~T_{\theta\theta}=T_{\varphi\varphi}/\sin^{2}\theta
=R^{2}P.
\end{equation}
The Einstein equations Eq.(1) then become
\begin{eqnarray}
\partial_{\pm}\partial_{\pm}R+\partial_{\pm}f\partial_{\pm}R
&=&-4\pi R\psi_{\pm}, \\
R\partial_{+}\partial_{-}R+\partial_{+}R\partial_{-}R
+{1\over 2}e^{-f}&=&4\pi R^{2}\Bigl(\varepsilon+{1\over 8\pi}
e^{-f}\Lambda\Bigr), \\
R \partial_{+} \partial_{-}f + 2\partial_{+} R \partial_{-} R
+e^{-f} &=& 8\pi R^{2}(\varepsilon +e^{-f}P).
\end{eqnarray}
By using the above equations, the derivatives of $M$
with respect to $\zeta_{\pm}$ are obtained as
\begin{equation}
\partial_{\pm}M=4\pi e^{f}R^{3}(\varepsilon\vartheta_{\pm}
-\psi_{\pm}\vartheta_{\mp}).
\end{equation}

Here we assume that the matter satisfies the dominant energy
condition, i.e., $\varepsilon \geq 0$ and $\psi_{\pm} \geq 0$.
Then, from Eq.(A7), we find that in the untrapped region
$\vartheta_{+}>0$ and $\vartheta_{-}<0$, the
quasi-local energy satisfies $\partial_{+}M \geq 0$ and
$\partial_{-}M \leq 0$.
If $z^{\mu}$ is outgoing spatial vector, the derivative
of $M$ along this direction is given by
\begin{equation}
z^{\mu}\partial_{\mu}M=\beta_{+}\partial_{+}M-\beta_{-}\partial_{-}M,
\end{equation}
where $\beta_{\pm}$ are positive. Hence, in the
untrapped region, $z^{\mu}\partial_{\mu}M \geq 0$ and
this means that $M$ is non-decreasing in any outgoing null
or spatial direction in this region.


\end{document}